\documentclass[twocolumn]{aastex6}

\begin{document}

\title{First results from the MADCASH Survey: A Faint Dwarf Galaxy Companion to the Low Mass Spiral Galaxy NGC 2403 at 3.2 Mpc \footnotemark[1]} \footnotetext[1]{Based in part on data collected at Subaru Telescope, which is operated by the National Astronomical Observatory of Japan.}

\author{Jeffrey L. Carlin\altaffilmark{2}}

\altaffiltext{2}{LSST and Steward Observatory, 933 North Cherry Avenue, Tucson, AZ 85721, USA; \\ Haverford College, Department of Astronomy, 370 Lancaster Avenue, Haverford, PA 19041, USA; jeffreylcarlin@gmail.com} 

\author{David J. Sand\altaffilmark{3}}
\altaffiltext{3}{Texas Tech University, Physics Department, Box 41051, Lubbock, TX 79409-1051, USA}

\author{Paul Price\altaffilmark{4}}
\altaffiltext{4}{Department of Astrophysical Sciences, Princeton University, Princeton, NJ 08544, USA}

\author{Beth Willman\altaffilmark{2}}

\author{Ananthan Karunakaran\altaffilmark{5}}
\altaffiltext{5}{Department of Physics, Engineering Physics and Astronomy, Queen's University, Kingston, Ontario, Canada, K7L 3N6}
\author{Kristine Spekkens\altaffilmark{5,6}}
\altaffiltext{6}{Department of Physics, Royal Military College of Canada, P.O. Box 17000, Station Forces, Kingston, ON K7L 7B4, Canada}

\author{Eric F. Bell\altaffilmark{7}}
\altaffiltext{7}{Department of Astronomy, University of Michigan, 1085 South University Ave, Ann Arbor, MI 48109, USA}

\author{Jean P. Brodie\altaffilmark{8}}
\altaffiltext{8}{University of California Observatories, 1156 High Street, Santa Cruz, CA 95064, USA}

\author{Denija Crnojevi{\'c}\altaffilmark{3}}

\author{Duncan A. Forbes\altaffilmark{9}}
\altaffiltext{9}{Centre for Astrophysics and Supercomputing, Swinburne University, Hawthorn VIC 3122, Australia}

\author{Jonathan Hargis\altaffilmark{10}}
\altaffiltext{10}{Space Telescope Science Institute, 3700 San Martin Drive, Baltimore, MD 21218, USA}

\author{Evan Kirby\altaffilmark{11}}
\altaffiltext{11}{California Institute of Technology, 1200 E. California Boulevard, MC 249-17, Pasadena, CA 91125, USA}

\author{Robert Lupton\altaffilmark{4}}

\author{Annika H. G. Peter\altaffilmark{12}}
\altaffiltext{12}{CCAPP, Department of Physics, and Department of Astronomy, The Ohio State University, Columbus, OH 43210, USA}

\author{Aaron J. Romanowsky\altaffilmark{8, 13}}
\altaffiltext{13}{Department of Physics and Astronomy, San Jos\'e State University, One Washington Square, San Jos\'e, CA 95192, USA}

\and

\author{Jay Strader\altaffilmark{14}}
\altaffiltext{14}{Department of Physics and Astronomy, Michigan State University, East Lansing, MI 48824, USA}

\begin{abstract}

We report the discovery of the faintest known dwarf galaxy satellite of an LMC stellar-mass host beyond the Local Group, based on deep imaging with Subaru/Hyper Suprime-Cam.  MADCASH~J074238+652501-dw lies $\sim$35~kpc in projection from NGC~2403, a dwarf spiral galaxy at $D$$\approx$3.2~Mpc. 
This new dwarf has $M_{g} = -7.4\pm0.4$ and a half-light radius of $168\pm70$~pc, at the calculated distance of $3.39\pm0.41$~Mpc. The color-magnitude diagram reveals no evidence of young stellar populations, suggesting that MADCASH~J074238+652501-dw is an old, metal-poor dwarf similar to low luminosity dwarfs in the Local Group.  The lack of either detected H\textsc{I} gas ($M_{\rm HI}/L_{V} < 0.69~M_\odot/L_\odot$, based on Green Bank Telescope observations) or {\it GALEX} NUV/FUV flux enhancement is consistent with a lack of young stars. 
This is the first result from the MADCASH (Magellanic Analog Dwarf Companions And Stellar Halos) survey, which is conducting a census of the stellar substructure and faint satellites in the halos of Local Volume LMC analogs via resolved stellar populations. Models predict a total of $\sim$4-10 satellites at least as massive as MADCASH~J074238+652501-dw
around a host with the mass of NGC~2403, with 2-3 within our field of view, slightly more than the one such satellite observed in our footprint.

\end{abstract}


\section{Introduction} \label{sec:intro}

The faint end of the galaxy luminosity function is an important probe of the astrophysics associated with the $\Lambda$+Cold Dark Matter ($\Lambda$CDM) model of galaxy formation. Quantitative verification of this model has met with challenges on sub-galactic scales (e.g., the ``missing satellites problem", \citealt{Klypin99}, and ``too big to fail", \citealt{Boylan12}), but significant progress has been made with the latest generation of numerical simulations, which include a wide range of baryonic physics \citep[e.g., ][]{Wetzel16}.  Likewise, the recent boom of faint dwarf discoveries in the Local Group \citep[LG; most recently][and references therein]{Torrealba16} has partially closed the gap between observations and theoretical expectations.  Many more systems should be discovered going forward \citep[e.g.,][]{Hargis14}.  Intriguingly, several of the newly discovered faint dwarfs may be associated with the Large Magellanic Cloud \citep[LMC;][among others]{bdb+15,kbt+15, jeb16, snk+16}, which is expected to have its own satellite system in the $\Lambda$CDM model \citep[e.g., ][]{Donghia08,snc+11}.

To comprehensively compare observations with expectations for galaxy formation in a $\Lambda$CDM universe, we must also look beyond the LG to measure the abundance and properties of dwarfs around primary galaxies of different masses, morphologies, and environments. This work has already begun for several systems with masses similar to, or greater than, the Milky Way (MW; e.g., M81: \citealt{cjt+13}; Cen~A: \citealt{csc+14, css+16}; NGC~253: \citealt{scs+14, rmm+16, tss+16}).
However, little attention has been paid to less-massive hosts \citep[but see][in NGC~3109]{ssc+15}, which may shed light on the putative dwarf galaxies of the LMC.

This paper presents the discovery of MADCASH~J074238+652501-dw, a low-luminosity satellite of the LMC stellar-mass analog NGC~2403 $(D \approx 3.2$~Mpc, $M_{\rm star} \sim 7\times10^9 M_\odot$, or $\sim2\times$ LMC stellar mass), the first result of a program to search for faint dwarfs and map the stellar halos of LMC analogs in the nearby Universe.
Prior to our discovery, NGC~2403 had one known satellite (DDO~44, $M_{B} \sim -12.1$; \citealt{kmk13}). 
In Section~\ref{sec:survey} we briefly summarize our survey plans and strategy to map the halos of nearby LMC-sized systems. Section~\ref{sec:data} discusses the observations and data reduction, and Section~\ref{sec:dwarf} details the properties of the newly discovered dwarf galaxy. We conclude in Section~\ref{sec:conclusions} by placing MADCASH~J074238+652501-dw in context both with respect to expectations from $\Lambda$CDM and with known systems in the Local Volume.

\section{Survey Description}\label{sec:survey}

We designed the MADCASH (Magellanic Analog Dwarf Companions And Stellar Halos) survey to use resolved stars to map the virial volumes of Local Volume galaxies ($d\lesssim$4~Mpc) with stellar masses of 1-7$\times10^9~M_\odot$ (roughly $1/3$ to 3 times that of the LMC, assuming $(M/L)_K = 1$).

 The \citet{kmk13} catalog includes four such galaxies -- NGC~2403, NGC~247, NGC~4214 and NGC~404 -- that are accessible from the Subaru telescope on Mauna Kea and have $E(B-V) < 0.15$. Two additional systems (NGC~55 and NGC~300; with $D<3$~Mpc) are in the southern sky and could be observed with the Dark Energy Camera (DECam). The virial radii of galaxies in this stellar mass range are $\sim$100-130 kpc, inferred from semi-analytic galaxy catalogs generated from the Millennium-WMAP7 structure formation model of \citet{gwa+13}\footnotemark[1]\footnotetext[1]{Searchable at \url{http://gavo.mpa-garching.mpg.de/MyMillennium/}.}.
At the time of this writing, the MADCASH team has acquired significant data (and upcoming observing time) on NGC~2403, NGC~247 and NGC~4214 (through NOAO Gemini-Subaru exchange time: PI Willman, 2016A-0920; and Keck-Subaru: PI Brodie, 2015B\_U085HSC, 2016B\_U138HSC).

\begin{figure}[!t]
\includegraphics[width=1.0\columnwidth, trim=7.0cm {0cm} {7.0cm} {0cm}, clip]{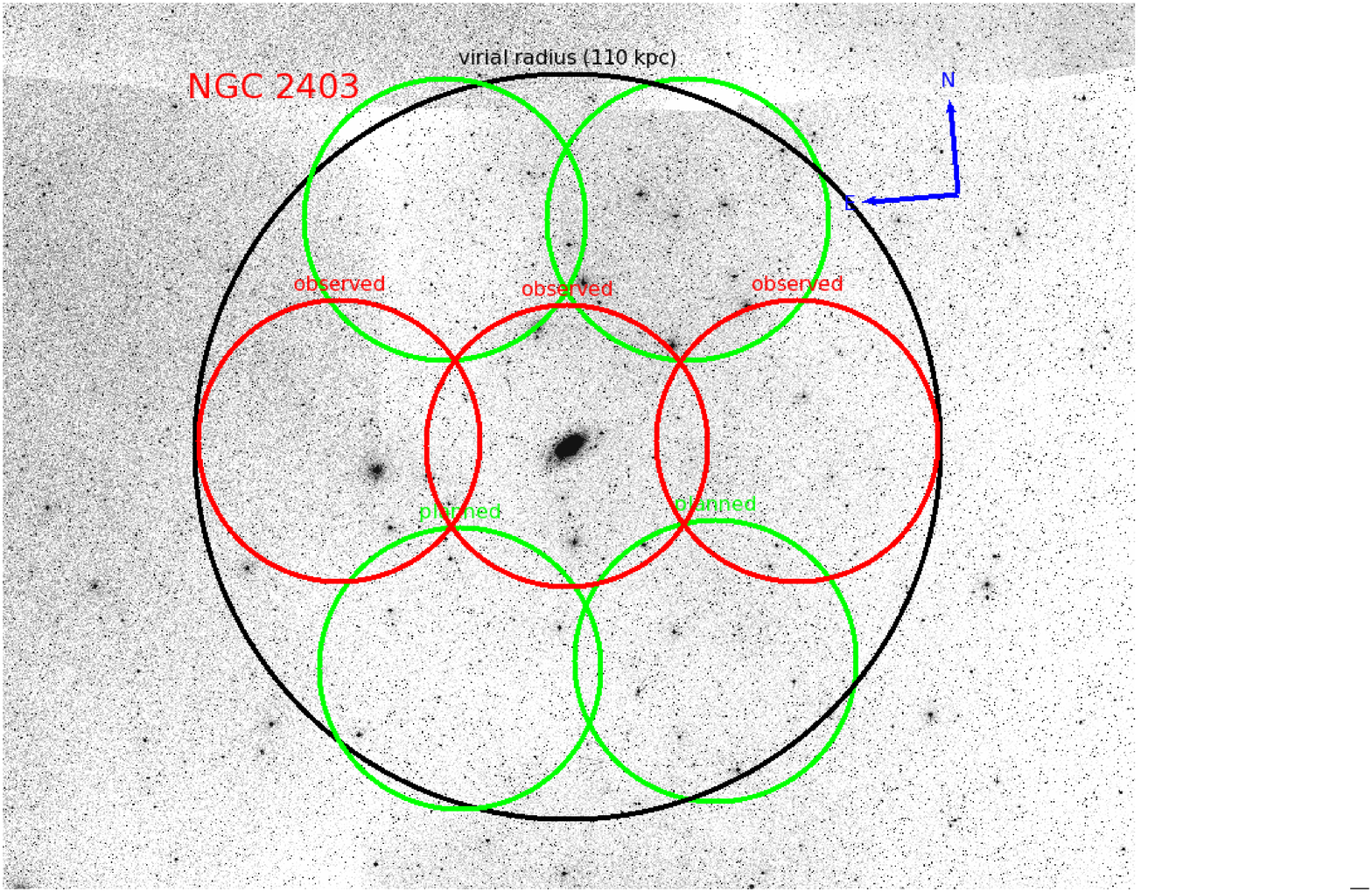}
\caption{DSS image of NGC~2403 with the HSC field of view overlaid as green and red circles. The large black circle represents an assumed virial radius of 110~kpc. Red circles are fields already observed, encompassing $\sim45\%$ of the virial volume of NGC~2403's halo. Green fields are the four additional HSC fields planned to complete our mapping of NGC~2403. }\label{fig:n2403_obs}
\end{figure}

The large aperture of the Subaru telescope and the 1.5$^{\circ}$ diameter field of view of the prime focus imager Hyper Suprime-Cam \citep[HSC;][]{Miyazaki12} make this project possible; the HSC field corresponds to $\sim$80 kpc at the distance to NGC~2403 ($D$$\sim$3.2 Mpc).  HSC can map the entire virial volume of an LMC-sized halo (see Figure~\ref{fig:n2403_obs}) in only seven pointings. We image our fields in $g$ and $i$ band to a depth that is $\sim$2 magnitudes below the tip of the red giant branch (TRGB). This is deep enough to identify and characterize dwarf galaxies as faint as $M_{V}\sim-7$.

\begin{figure*}[!t]
\includegraphics[width=1.0\textwidth]{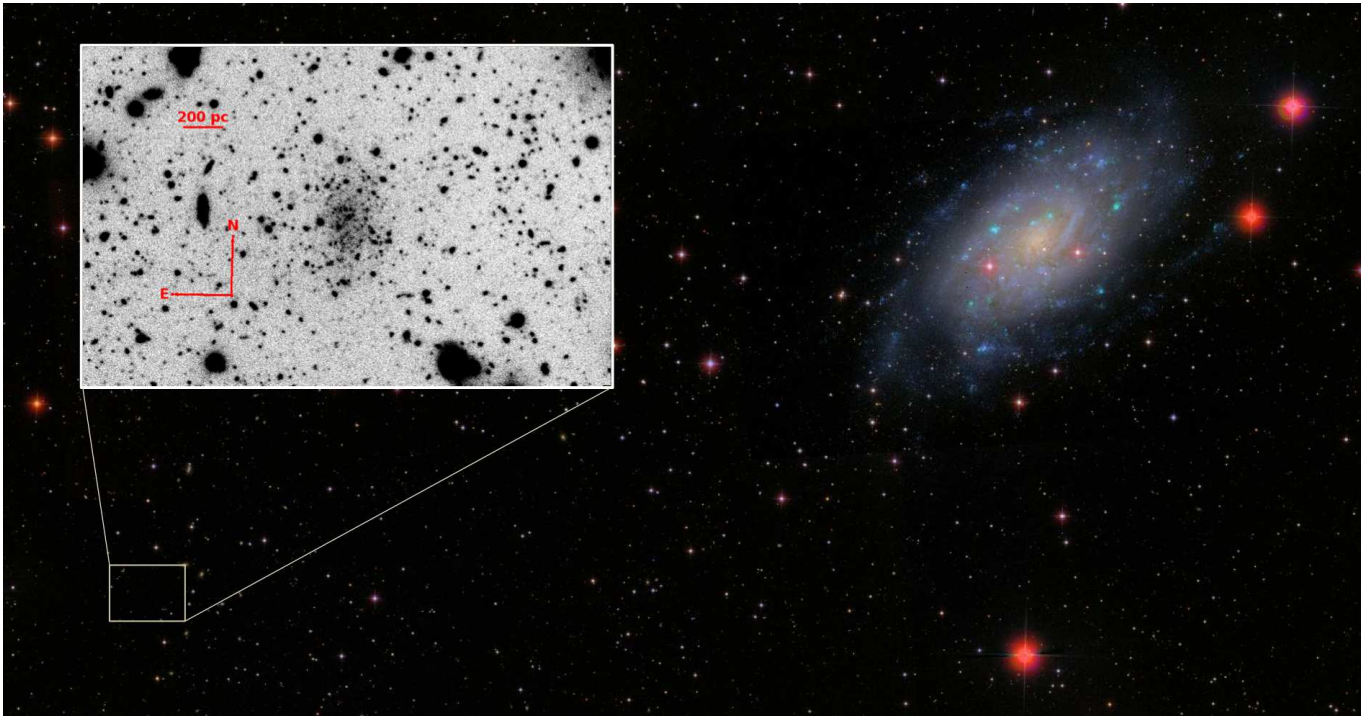}
\caption{{\it Inset:} HSC image (in $g$-band, $\sim2.8' \times 1.8'$ in size) of the candidate dwarf galaxy MADCASH~J074238+652501-dw.  
The center of NGC~2403 is $\sim38'$ away, or $\sim35$~kpc in projection. {\it Background image:} Color composite from SDSS-III (acquired via \url{http://wikisky.org/}) of NGC~2403. }\label{fig:n2403_dwarf}
\end{figure*}

\section{Observations and Data Reduction} \label{sec:data}

We observed three NGC~2403 fields with Subaru/HSC on 2016 Feb 9-10 with exposure times of $10\times300$~s in $g$ and $10\times120$~s in $i$ for each field.  The seeing was $\sim$$0.5-0.7''$. 
The fields, shown in Figure~\ref{fig:n2403_obs}, extend nearly to the virial radius both to the east and west of the main body of NGC~2403 and sample $\sim45\%$ of NGC~2403's virial volume.

The images were processed using the HSC pipeline (hscPipe~4.0.1; \url{http://hsca.ipmu.jp}; \url{http://hsc.mtk.nao.ac.jp/pipedoc_e/index.html}), which is based on an earlier version of the LSST pipeline \citep{akl+10}.
Images are bias-subtracted, flat-fielded with dome flats, corrected for the brighter-fatter effect (Coulton et al., in prep.), and astrometrically and photometrically calibrated against Pan-STARRS~1 Processing Version 2 (\citealt{sfj+12,tsl+12,msf+13}).
The images are transformed to a common reference frame and coadded with conservative clipping to remove artifacts that appear on a single visit. Coadded images are used for all the photometric and astrometric measurements reported below.  
We create a merged source list from deblended catalogs in each band, and apply the same centroid and aperture or model (generally derived from the $i$-band image) to measure each object's flux in both bands.
Point sources are separated from extended sources by removing objects for which the model and PSF fluxes differ by more than three times the flux error for that object. All stellar magnitudes presented in this work are derived from PSF photometry.
We then match our catalogs to SDSS DR9 \citep{Ahn12}, and transform to $g_{\rm SDSS}$ and $i_{\rm SDSS}$ magnitudes with an offset and color term. The transformed magnitudes have $\sim0.03$~mag scatter about the SDSS values. All magnitudes presented henceforth are on the SDSS photometric system, corrected for extinction using the \citet{sfd98} maps with coefficients from \citet{sf11}. The average color excess for stars in this region of the sky is $E(B-V)\sim0.048$.

To estimate the photometric completeness of our catalogs, we match our Subaru/HSC data to three {\it HST/ACS} fields in the halo of NGC~2403 from the GHOSTS program \citep{rds+11}\footnotemark[2]\footnotetext[2]{Data products available at \url{https://archive.stsci.edu/pub/hlsp/ghosts/index.html}}.  The GHOSTS fields were observed with the F606W and F814W filters. Artificial star tests showed that the ACS data are $>90\%$ complete to the magnitude limit of our HSC photometry \citep{rds+11}. Using a matching radius of $1''$, we recover half of the HST/ACS stellar sources (i.e., we are 50\% complete) at a magnitude of F814W$\approx26.0$, which corresponds to $i \approx 26.4$ in our HSC data.

\begin{figure*}[!t]
\begin{center}
\includegraphics[width=0.9\textwidth, trim=0.5cm {0.0cm} {0.5cm} {0.5cm}, clip]{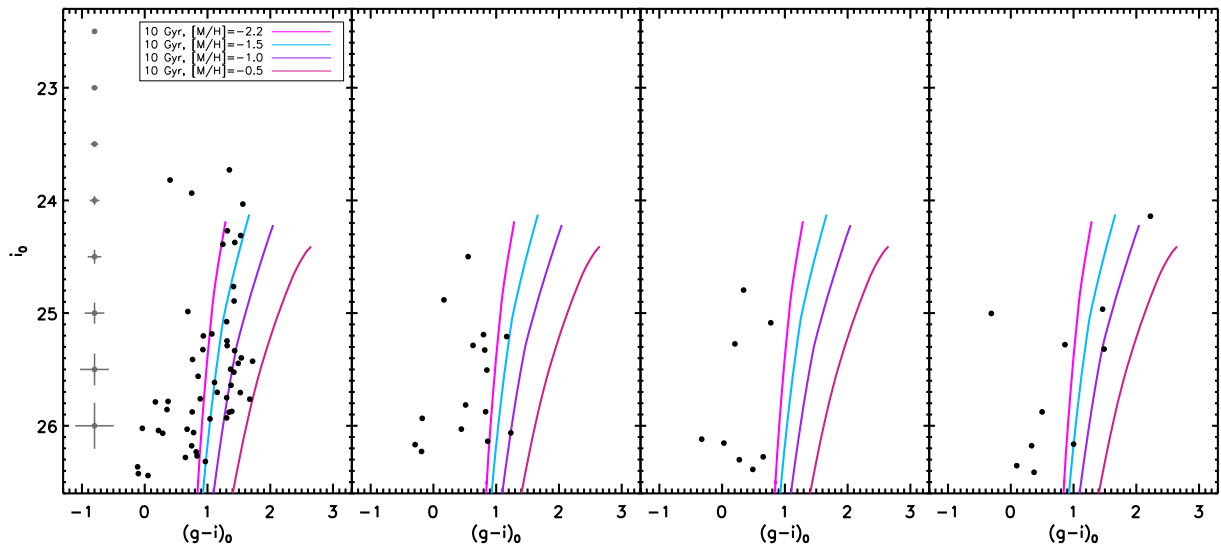}
\caption{{\it Left panel:} Color magnitude diagram of point sources within $20''$ of
 MADCASH~J074238+652501-dw. Typical photometric errors as a function of magnitude are shown at the left side of this panel. The other three panels show CMDs of nearby ``blank-sky'' regions of the same size. In each panel, we show Padova isochrones of old (10-Gyr) populations with metallicities of [M/H]=$-$2.2, $-$1.5, $-$1.0, and $-$0.5, at $D=3.39$~Mpc (the distance to MADCASH~J074238+652501-dw; see Section~\ref{subsec:dist}). 
There is an overdensity of resolved sources at the position of the dwarf, consistent with an old, metal-poor RGB, 
extending $\sim$2 mags below the RGB tip. 
}
\label{fig:n2403_dwarfCMD}
\end{center}
\end{figure*}

\section{A New Dwarf Galaxy Companion} \label{sec:dwarf}

We visually identified a candidate dwarf galaxy $\sim$35 kpc to the east of NGC~2403 in projection (Figure~\ref{fig:n2403_dwarf}), which we dub MADCASH~J074238+652501-dw.  At this radius, MADCASH~J074238+652501-dw is just beyond the field of view of previous work on the halo of NGC~2403 \citep{Barker12}. This dwarf shows no sign of a nuclear star cluster or disturbed morphology. NGC~2403 has one other known dwarf companion, the bright dSph/dE galaxy DDO~44 ($M_{B} = -12.1$; \citealt{kmk13}) roughly $1.3^\circ$ ($\sim73$~kpc in projection) to the north.

The left panel of Fig.~\ref{fig:n2403_dwarfCMD} shows a color--magnitude diagram of point sources within $20''$ of the center of MADCASH~J074238+652501-dw. For comparison, other panels show CMDs in randomly selected nearby background fields (also with $20''$ radius). Each panel contains PARSEC isochrones \citep{bmg+12} for old (10 Gyr) populations at our derived distance to MADCASH~J074238+652501-dw (3.39~Mpc; see Sec.~\ref{subsec:dist}) and metallicities ([M/H]) of -2.2, -1.5, -1.0, and -0.5 (assuming a solar metallicity of $Z_\odot = 0.0152$). The most metal-poor isochrones follow the red giant branch (RGB) of MADCASH~J074238+652501-dw closely, with little evidence of younger populations blueward of the RGB or more metal-rich RGB stars following the reddest of the isochrones. 
 Two conclusions can be drawn from Figure~\ref{fig:n2403_dwarfCMD} -- first, that there is an obvious stellar excess relative to neighboring regions, and secondly that the excess stars predominantly cluster around the old, metal-poor (${\rm [M/H]} < -1.0$) locus of the isochrones we have overlaid. 
Assuming that the stars in MADCASH~J074238+652501-dw mostly cluster around [Fe/H] = -2, then we conclude that the galaxy does not host populations significantly younger than 10 Gyr.

We estimate the mean metallicity of MADCASH~J074238+652501-dw by comparing the CMD positions of stars with PARSEC isochrones. In order to eliminate contamination by non-members, we select the 13 stars in the left panel of Fig.~\ref{fig:n2403_dwarfCMD} with $i_0 < 25$ and $(g-i)_0 > 1.0$. We linearly interpolate 10 and 14~Gyr isochrones shifted to the distance modulus listed in Table~\ref{tab:params}, and assign each star the metallicity of the interpolated isochrone that passes through its color and magnitude. The mean metallicity of the 13 stars is ${\rm [Fe/H]} = -1.6 (-1.7)$ for the 10 (14) Gyr isochrones, with standard deviation of 0.4~dex.

\subsection{Distance}\label{subsec:dist}

We estimate the distance to MADCASH~J074238+652501-dw using the TRGB method \citep{lfm93}. The TRGB absolute magnitude is estimated by averaging the magnitudes of the brightest metal-poor (${\rm [M/H]} = -2.2, -1.5, {\rm and} -1.0$), old (10 Gyr) stars in the PARSEC isochrones; we adopt a TRGB magnitude of $M_i^{\rm TRGB} = -3.47\pm0.05$.
We locate the TRGB of MADCASH~J074238+652501-dw using stars within $20''$ of the dwarf center, keeping only stars with $0.8 < (g-i)_0 < 2.1$ (the color range of the metal-poor RGB; see Fig.~\ref{fig:n2403_dwarfCMD}). We bin these in magnitude to create a luminosity function, then use a zero-sum Sobel edge-detection filter to locate the transition in stellar density corresponding to the RGB tip. We repeat this for bins of different widths (from 0.15 to 0.2~mag in steps of 0.01) and then adopt the mean value; the standard deviation is taken as the uncertainty.\footnotemark[3]\footnotetext[3]{We attempted to apply the maximum likelihood method of \citet{mmr+06}, but it failed to converge due to the small number of resolved stars near the TRGB of the dwarf.} We measure $i_{0, {\rm TRGB}} = 24.18\pm0.20$ for MADCASH~J074238+652501-dw \citep[in agreement with the distance to NGC~2403 from previous work; e.g.,][]{Bellazzini08}, corresponding to $m-M = 27.65\pm0.26$ ($D = 3.39\pm0.41$ Mpc) for the new dwarf galaxy. For comparison, we also perform the TRGB analysis on stars near the main body of NGC~2403, and find $i_{0, {\rm TRGB}} = 23.92\pm0.11$ for NGC~2403, or a distance modulus of $m-M = 27.39\pm0.16$ ($D = 3.01\pm0.23$ Mpc). This agrees with typical measurements of the distance to NGC~2403 within the uncertainties (e.g., \citealt{rds+11}: $m-M = 27.51\pm0.07$) .


\subsection{Structural parameters}\label{subsec:params}

To derive structural parameters of MADCASH~J074238+652501-dw, we select RGB candidates centered on the three most metal-poor isochrones in Figure~\ref{fig:n2403_dwarfCMD}, with magnitudes $23.7 < i_0 < 26.2$, within a $5'\times5'$ box centered on the dwarf. This catalog was passed to a maximum likelihood estimator of the dwarf structural parameters using the \citet{Sand12} implementation of the \citet{mdr08} method. The resulting structural parameters for MADCASH~J074238+652501-dw are given in Table~\ref{tab:params}, with uncertainties determined via 1000 bootstrap resamplings of the data.

The central position of MADCASH~J074238+652501-dw is well constrained,
but the additional parameters are poorly measured due to the small number of resolved stars in the dwarf. We derive a half-light radius of $r_{\rm h} = 10.2\pm3.0''$, corresponding to 168~pc at a distance of 3.39~Mpc. For the ellipticity, we derive only an upper limit of $\epsilon < 0.42$ (within 68\% confidence limits). Because the ellipticity is poorly constrained, we cannot reliably measure the dwarf's position angle; the value of $\theta = 29^\circ$ corresponds to the likelihood maximum, but is unconstrained.

\subsection{Luminosity and stellar populations}\label{subsec:lumin}

\begin{figure}[!t]
\includegraphics[width=1.0\columnwidth, trim=1.0cm {0.0cm} {0.0cm} {0.0cm}, clip]{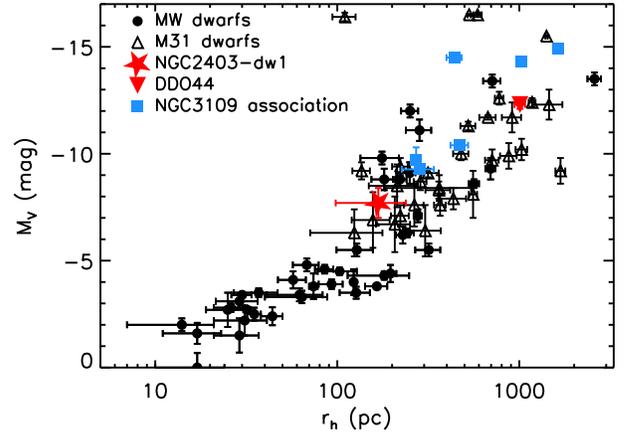}
\caption{Absolute $V$-band magnitude ($M_{V}$) and half-light radius ($r_{\rm h}$) of MADCASH~J074238+652501-dw (large red star) placed in context with other known systems. Black points and triangles are data for MW and M31 dwarf galaxies \citep[e.g.,][]{m12},
and blue squares are members of the NGC~3109 association, including Antlia and Antlia B \citep{ssc+15}. The small, red, downward-pointing triangle represents DDO~44, another known satellite of NGC~2403 \citep{ksg+99}.}
\label{fig:newdwarf_mv_rh}
\end{figure}

We estimate the luminosity of MADCASH~J074238+652501-dw by measuring the integrated flux in a circular aperture 
of radius $r_{\rm h} = 10.2''$ and central position as measured in Section~\ref{subsec:params} (see, e.g., \citealt{scs+14, ssc+15}). 
After subtracting the average background from 100 equal-area apertures in random positions throughout the same CCD frame and multiplying the flux by two to account for our half-light radius aperture, we find integrated luminosities of $M_{g} = -7.4\pm0.4$ and $M_{i} = -8.1\pm0.6$.
This translates to $M_{V} = -7.7\pm0.7$ using the filter transformations of \citet{jga06}.

The measured luminosity and half-light radius place MADCASH~J074238+652501-dw upon the observed relationship for LG dwarfs (Figure~\ref{fig:newdwarf_mv_rh}).
This suggests that even though MADCASH~J074238+652501-dw has evolved in a different environment than MW/M31 dwarfs, the physical processes that determine its properties are similar to those in more massive hosts' halos. If MADCASH~J074238+652501-dw also follows other scaling relations for LG dwarf galaxies \citet[e.g.,]{m12}, then this dwarf with $M_{V} = -7.7$ should have a dynamical mass-to-light ratio of $\sim100$, dynamical mass of $M_{\rm dyn} (< r_{\rm h}) \sim 5\times10^6~M_\odot$, and a mean metallicity of [Fe/H]$\sim-2.0$. This metallicity is consistent with our inference of an overall metal-poor population in MADCASH~J074238+652501-dw.

The sparse CMD precludes a precise quantification of the SFH beyond our previous statement that the galaxy does not host stars significantly younger than 10 Gyr.

\subsection{GALEX UV}

We discern no FUV or NUV flux enhancement in available {\it GALEX} imaging at the position of MADCASH~J074238+652501-dw. The nearest source in the MAST/{\it GALEX} GR6 catalog, at $\sim0.86'$ from the center of the dwarf, is faint in the FUV, with no NUV detection. This source has no obvious counterpart in our HSC images. 

 
Using tiles from the {\it GALEX} All-Sky Imaging Survey, we measure the FUV flux of MADCASH~J074238+652501-dw, using the same aperture and background-subtraction technique used in estimating the optical luminosity in Section~\ref{subsec:lumin}. This flux was converted into a luminosity using our adopted TRGB distance modulus of 27.65, adopting an extinction coefficient $A_{\rm FUV} = 7.9~E(B-V)$ \citep{lgt+09}. We convert this to a star formation rate (SFR) using the relation between SFR and UV continuum from \citet{msd+15}: ${\rm SFR} (M_\odot~{\rm yr}^{-1}) = 2.04\times10^{-28}~L_\nu ({\rm erg~s}^{-1}~{\rm Hz}^{-1})$, and find an upper limit on the SFR in MADCASH~J074238+652501-dw of $4.4\times10^{-6}~M_\odot~{\rm yr}^{-1}$. The same relation using the {\it GALEX} FUV aperture magnitude for MW dwarf Leo~T from \citet{lgk+11} yields SFR$_{\rm Leo T} \approx 5.6\times10^{-6}~M_\odot~{\rm yr}^{-1}$, suggesting that MADCASH~J074238+652501-dw is at most forming stars at a similar rate as Leo~T.

\begin{deluxetable}{lr}
\tablecaption{Properties of MADCASH~J074238+652501-dw \label{tab:params}}
\tablehead{\colhead{Parameter} & \colhead{Value}}
\startdata
RA (hh:mm:ss) & $07:42:38.887  \pm 2.6''$ \\
Decl (dd:mm:ss) & $+65:25:01.89 \pm 3.2''$ \\
$m-M$ (mag) & $27.65 \pm 0.26$ \\
$D$ (Mpc) & $3.39 \pm 0.41$ \\
$M_{g}$ (mag) & $-7.4 \pm 0.4$ \\
$M_{V}$ (mag) & $-7.7 \pm 0.7$ \\
$r_{\rm h}$ (arcsec) & $10.2 \pm 3.0$ \\
$r_{\rm h}$ (pc) & $168 \pm 70$ \\
$\epsilon$ & $<0.42~(68\%~{\rm CL})$ \\
$\theta$ (deg.) & $29^\circ ({\rm unconstrained})$ \\
$\mu_0$ (mag arcsec$^{-2}$) & $25.9 \pm 0.7$\\
$M_{\rm star} (M_\odot$) & $\sim1\times10^5$ \\
$M_{\rm HI}/L_{V} (M_\odot/L_\odot)$ & $<0.69$ \\
$M_{\rm HI} (M_\odot)$ & $<7.1\times10^4$ \\
SFR ($M_\odot$/yr)& $\leq 4.4\times10^{-6}$
\enddata
\end{deluxetable}

\subsection{GBT H\textsc{I} observations}

Using the Robert C. Byrd Green Bank Telescope\footnote[4]{The National Radio Astronomy Observatory is a facility of the National Science Foundation operated under cooperative agreement by Associated Universities, Inc.}, we obtained position-switched H\textsc{I} observations of MADCASH~J074238+652501-dw through Director's Discretionary Time (AGBT-16A-462; PI:Spekkens) on 2016 May 9 and 11. The GBT spectrum in the velocity ranges $-1000 \leq V_{\rm LSRK} \leq -50$~km~s$^{-1}$ and $50 \leq V_{\rm LSRK} \leq 1000$~km~s$^{-1}$ has rms noise of $\sigma = 0.35$~mJy at a spectral resolution of 15~km~s$^{-1}$. We do not find any H\textsc{I} emission in these velocity ranges within the FWHM = $9.1' (8.97$~kpc) GBT beam at this frequency. This non-detection combined with the measured distance and luminosity suggest that a putative H\textsc{I} counterpart has a $5\sigma$, 15~km~s$^{-1}$ H\textsc{I} mass upper limit of $M_{\rm HI, lim} = 7.1\times10^4 M_\odot$ and $M_{\rm HI} / L_{V} = 0.69~M_\odot / L_\odot$. The satellite is therefore gas-poor, similar to what is found for other dwarf spheroidals in the Local Volume \citep{gp09, sum+14}.

\section{Conclusions} \label{sec:conclusions}

We report the discovery of a faint ($M_{g} = -7.4\pm0.4$) dwarf galaxy companion of the LMC analog NGC~2403 ($\sim2\times$ LMC stellar mass) in imaging data from the MADCASH survey using Hyper Suprime-Cam on the Subaru telescope. From resolved stars reaching $\sim2$~magnitudes below the RGB tip, we show that the new dwarf, MADCASH~J074238+652501-dw, has predominantly old, metal-poor stellar populations ($\sim10$~Gyr, ${\rm [M/H]} \sim -2$) similar to those in the LG ultra-faint galaxies. 
Using non-detections in H\textsc{I} and UV observations, we place upper limits on the available gas reservoir and star formation rate, confirming this as a gas-poor system with old stellar populations.
Our derived distance modulus of $m-M = 27.65\pm0.26$ places MADCASH~J074238+652501-dw near NGC~2403, bolstering the case that it is a faint satellite of this Local Volume LMC analog. 

Should we have expected to find only one new satellite? NGC~2403 has one previously known satellite, the dwarf galaxy DDO~44\footnotemark[5]\footnotetext[5]{\citet{kdg+02} suggested NGC~2366, Holmberg~II, UGC~4483, and KDG052 may also be companions of NGC~2403. Of these, only NGC~2366 ($M_{B} = -16.1$) has NGC~2403 designated as its main tidal disturber by \citealt{kkm14}; all others are predominantly influenced by M81. 
For this work, we assume that these galaxies are not satellites of NGC~2403.}
($M_{V} \sim -12.5, {\rm [Fe/H]} \sim -1.7$; \citealt{ksg+99}), which is outside our current footprint. The stellar 
masses of this satellite and the newly discovered dwarf are $M_{\rm star, DDO~44} \sim 6\times10^7~M_\odot$ (estimated from the $K$-band 
magnitude from \citealt{kmk13}, assuming $M/L = 1$) and $M_{\rm star} \sim 1\times10^5~M_\odot$ for 
MADCASH~J074238+652501-dw.  Based on abundance-matching relations \citep{mnw13,gbb+14}, 
applied to subhalo mass functions from dark-matter-only simulations (e.g., \citealt{gbb+14}), we expect 3-11 
dwarf galaxies at least as massive as MADCASH~J074238+652501-dw in NGC 2403's virial volume.  If satellites follow the subhalo 
distribution in dark-matter-only simulations \citep{hcf+16}, which is nearly isothermal and consistent with the distribution 
of satellites at high redshift \citep{nat+11}, we should have found 2-3 satellites in our footprint with $M_{\rm star} 
\gtrsim 1\times10^5~M_\odot$.  This suggests that we should find a 
factor of 2-3 more in a complete survey of the remaining $\sim55\%$ of NGC~2403's virial volume. While this simple estimate suggests that one dwarf galaxy is fewer than we expect to find in our current data, we must observe the entire virial volume of NGC~2403 to fully assess the significance of its satellite abundance. Placing definitive constraints on cosmological models will require mapping the virial halos of the ensemble of hosts, which sample a variety of environments, in our MADCASH program.

\acknowledgments

We thank Fumiaki Nakata and Rita Morris for assistance at the Subaru Telescope, Mike Beasley for attempting to obtain a spectrum of the new dwarf, the referee for helpful comments, and Michael Wood-Vasey for conversations that helped improve our photometry. JLC and BW acknowledge support by NSF Faculty Early Career Development (CAREER) award AST-1151462.
DJS acknowledges support from NSF grant AST-1412504. The work of DJS was performed at the Aspen Center for Physics, which is supported by NSF grant PHY-1066293. JPB and AR are supported by NSF grant AST-1211995.
Some data presented here were obtained from the Mikulski Archive for Space Telescopes (MAST). STScI is operated by the Association of Universities for Research in Astronomy, Inc., under NASA contract NAS5-26555. Support for MAST for non-{\it HST} data is provided by the NASA Office of Space Science via grant NNX09AF08G and by other grants and contracts.

The Pan-STARRS1 Surveys have been made possible through contributions of the Institute for Astronomy, the University of Hawaii, the Pan-STARRS Project Office, the Max-Planck Society and its participating institutes, the Max Planck Institute for Astronomy, Heidelberg and the Max Planck Institute for Extraterrestrial Physics, Garching, The Johns Hopkins University, Durham University, the University of Edinburgh, Queen's University Belfast, the Harvard-Smithsonian Center for Astrophysics, the Las Cumbres Observatory Global Telescope Network Incorporated, the National Central University of Taiwan, the Space Telescope Science Institute, the National Aeronautics and Space Administration under Grant No. NNX08AR22G issued through the Planetary Science Division of the NASA Science Mission Directorate, the National Science Foundation under Grant AST-1238877, the University of Maryland, Eotvos Lorand University (ELTE), and the Los Alamos National Laboratory.



\vspace{5mm}
\facilities{Subaru (Hyper Suprime-Cam), GALEX, GBT, HST (ACS)}



\end{document}